\documentclass[12pt]{article}
\usepackage{graphicx}
\usepackage{makeidx}
\usepackage{amsmath}
\usepackage{amsfonts}
\usepackage{mathrsfs}
\usepackage{wasysym}
\usepackage{pstricks,pst-node,pst-coil,pst-tree}
\title{Mapping a Homopolymer onto a Model Fluid}
\author{S. Pasquali, J.K. Percus}

\newcommand{\beq}{\begin{equation}}
\newcommand{\eeq}{\end{equation}}
\newcommand{\beqa}{\begin{eqnarray}}
\newcommand{\eeqa}{\end{eqnarray}}
\renewcommand{\a}{\alpha}
\renewcommand{\b}{\beta}

\renewcommand{\d}{\delta}

\newcommand{\z}{\zeta}

\renewcommand{\l}{\lambda}

\renewcommand{\>}{\rangle}
\newcommand{\<}{\langle}

\renewcommand{\inf}{\infty}

\begin{document}
\begin{titlepage}
\hfill cond-mat/0604045
\vspace{20pt}

\begin{center}
{\large\bf{Mapping a Homopolymer onto a Model Fluid }}
\end{center}

\vspace{6pt}

\begin{center}
{\large S. Pasquali$^1$, J.K. Percus$^2$} \vspace{20pt}

{\em $^1$ PCT - ESPCI, 10 rue Vauquelin, 75231 Paris cedex 05, France\\
$^2$ Courant Institute of Mathematical Sciences and Department of Physics,\\ 
251 Mercer Street, New York NY 10012, USA}

\end{center}

\begin{center}
\textbf{Abstract}
\end{center}
\begin{quotation}\noindent
We  describe a linear homopolymer using a Grand Canonical
ensemble formalism, a statistical representation that is very convenient for formal
manipulations.
We investigate the properties of a system where only next neighbor
interactions and an external, confining, field are present, and then show how
a general pair interaction can be introduced perturbatively, making use of a
Mayer expansion.
Through a diagrammatic analysis, we shall show how constitutive equations derived
for the polymeric system are equivalent to the Ornstein-Zernike and
P.Y. equations for a simple fluid, and find the implications of such a mapping
for the simple situation of Van der Waals mean field model for the fluid.
\end{quotation}

\vfill
 \hrule width 5.cm
\vskip 2.mm
{\small
\noindent $^1$ e-mail: samuela@turner.pct.espci.fr\\
 $^2$ e-mail: jerome.percus@nyu.edu}
\end{titlepage}

\newpage
\section{Introduction}
Molecules with linear chains as descriptives backbones can, in whole or in part, be constructed by 
temporally ordered sequences of additions.
The efficiency of such a process recommends chains as major actors in molecular biology, as well as in 
industrial chemistry.
They may be examined at multiple levels of resolution.
In the study about to be reported, a very coarse description is used in which the units are taken as point
entities with next neighbor strong binding forces and weak non-next neighbor interactions, all in the 
presence of a confining external field.
Even more narrowly, we restrict our attention to homopolymers, but will indicate how questions concerning
the statistics of heteropolymers can be addressed within this format.

In further detail, we study a single homopolymer in classical thermal equilibrium.
The homopolymer context suggests, in analogy with fluids, extension to a bath of monomers which can join or
leave the chain, inducing fluctuations in the monomer number $N$ and creating what might be termed a 
monomer grand ensemble.
The corresponding formalism also follows the fluid model: suppose there are $N'$ monomers in the full system, $N$
of which are bound together to create a polymer, $N'-N$ being free in the volume V envisioned.
Then the canonical partition function of the complete system takes the form (one needs at least $N=1$ to 
declare a polymer)
\beq  \label{partition_canonical}
Q_{N',V}^{TOT} = \sum_{N \geq 1} \frac{\left(V Q^{(m)}\right)^{N'-N}}{\left(N'-N\right)!} Q_N^{(p)}(u)
\eeq
where $Q^{(m)}$ is the monomer canonical partition function in its center-of-mass coordinate system,
$Q^{(p)}$ that of the $N$-monomer polymer in the external potential field $u$.
The ``grand'' partition function of the monomer collection defining the polymer will then be taken as
\beq 
\Xi^{(p)}(\z)=\lim_{N',V \to \inf} \frac{N'!}{\left(V Q^{(m)}\right)^{N'}}\; Q_{N',V}^{TOT}
\eeq
or via the Stirling approximation,
\beqa \label{grand_partition}
\Xi^{(p)}(\z)&=&\lim_{N',N \to \inf} \sum_{N \geq 1}\left(\frac{N'/V}{Q^{(m)}}\right)^N Q_N^{(p)}(u) 
\nonumber \\
&&= \sum_{N \geq 1} \z_0^N Q_N^{(p)}(u), \;\;\;\;\;\; \z_0=\frac{\left(N'/V\right)}{Q^{(m)}},
\eeqa
identifying the fugacity as
\beq
\z(r) = \z_0 e^{-\b u(r)} = e^{\b\mu(r)}.
\eeq
The expression (\ref{grand_partition}) is very convenient for formal manipulations, but a certain amount of 
caution must be exercised.
In particular, number fluctuations can be very large, as we will see, so that ``canonical'' - i.e. fixed monomer
number - information, requires care.

In this paper, our aim is to show how (\ref{grand_partition}) allows for a familiar type of diagrammatic 
representation, which not only gives rise to a systematic perturbation expansion in the strength of the 
non-next neighbor forces, but also - under special but physically reasonable restrictions - maps onto a suitable
classical fluid, with its array of well-tested computational recipes.
With this, we can routinely examine the effect of structured fields on the polymer density profile, which 
is our implicit objective.

\section{The Reference System {\small \cite{Frisch}}}
Now let us introduce explicit notation following \cite{Percus}.
The model homopolymer that we will deal with has as monomer units point ``particles'' located at 
$\{r_i,i=1,\ldots,N\}$, which may be imagined as spatial locations, but can include type, internal state, etc.
There is an external potential field $u(r)$ appearing only in the combination $\z(r)=\z_0\; e^{-\b u(r)}$, where
$\b$ is the reciprocal temperature, and a next neighbor potential $\phi(r,r')$, typically $\phi(r-r')$, with associated
Boltzmann factor $w(r,r')=e^{-\b \phi(r,r')}$ as well as a general pair interaction (including next neighbors) 
$v(r,r')$, giving rise to the Mayer function $f(r,r')=e^{-\b v(r,r')}-1$.
Note that we can expect that $w \to 0, f\to 0$ as $|r-r'|\to \inf$.
When $r$ represents real space location alone, we have the format of freely
jointed units, but that is not intrinsic, since the units can involve more
than one monomer (see e.g. \cite{Muller}).

We will choose, as generating function for the thermodynamics, the grand canonical potential
\beq
\Omega[\z]=-(1/\b)\ln{\Xi^{(p)}[\z]},
\eeq
so that , e.g. the resulting monomer density profile will be given by
\beqa \label{density}
n(r) &=& -\z(r) \frac{\d\b\Omega[\z]}{\d \z(r)} \nonumber \\
&=& + \frac{\d\ln{\Xi^{(p)}[\z]}}{\d\ln \z(r)}
\eeqa

To start with, consider the ``reference system'' in which the general pair
interactions have not yet been included. 
Since the monomers can be labeled ordinally along the chain, and they are not
equivalent with respect to system energy, there is no statistical weight
factor for an $N$-monomer configuration.
Hence (\ref{grand_partition}) can be written as
\beqa \label{reference_parition}
\Xi^{(p)}[\z] &=& \z_0 \int e^{-\b u(r_1)}dr_1 + \z_0^2 \int
\int e^{-\b u(r_1)} w(r_1,r_2) e^{-\b u(r_2)} dr_1 dr_2 + \nonumber \\
&& +\; \z_0^3 \int\int\int  e^{-\b u(r_1)} w(r_1,r_2) e^{-\b u(r_2)}w(r_2,r_3)
e^{-\b u(r_3)} dr_1 dr_2 dr_3 + \ldots \nonumber \\
&=& \int \z(r) dr + \int\int \z\,w\,\z(r,r')+\z\,w\,\z\,w\,\z(r,r')+\ldots dr
dr' \nonumber \\
&=& \<1|\z\left(I-w\,\z\right)^{-1}|1\>
\eeqa
where $w$ is regarded as a (continuous  as needed) matrix, $\z$ a
diagonal matrix, $I$ ($I(r,r')=\d(r,r')$) the identity, and $1$ the vector of
all $1$'s.
The validity of (\ref{reference_parition}) is subject to convergence.
In terms of the molecular weight control parameter $\z_0$, it is apparent that
(\ref{reference_parition}) will converge only until $\z_{0\;MAX} = 1/\l_{MAX}(w
e^{-\b u})$ - the value at which the minimum eigenvalue of $w\,\z$ reaches
$1$.
The behavior of the system as $\z_0 \to \z_{0\;MAX}$ must be factored
out.
Combining (\ref{density}) and (\ref{reference_parition}), we have, using the
standard $(A^{-1})' = -A^{-1} A' A^{-1}$ for any derivative $(')$ on any
matrix inverse, the mean density profile
\beq
n(r) = \frac{\<1|\z\left(I-\z\,w\right)^{-1}|r\> \z(r) \<r|\left(I-w\,\z\right)^{-1}|1\>}{\Xi^{(p)}[\z]}
\eeq
In particular, the mean number of monomers, $\overline{N} = \int n(r) dr$,
becomes
\beq
\overline{N} = \frac{\<1|\z\left(I - w\,\z\right)^{-2}|1\>}{\<1|\z\left(I - w\,\z\right)^{-1}|1\>}
\eeq
dominated by $\overline{N}=\z_{0\;MAX}/(\z_{0\;MAX}-\z_0)$ as
$\z_0\to\z_{MAX}$.
Thus, large mean molecular weight occurs very close to the ``resonance'' at
$\z=\z_{0\;MAX}$.
The effect is brought out forcefully by computing the standard deviation of
$N$ in a similar fashion:
\beq
\d N = \left(\z_0 \frac{\d\overline{N}}{\d\z_0}\right)^{1/2}\qquad \sim \overline{N},
\eeq
evidence of something like an exponential distribution.

It is also useful to look ahead and see how the qualitative behavior will be
modified in the presence of non-next neighbor interactions, which of course
contribute $N^2$ terms to the configurational energy.
If all $N^2$ were positive and of similar magnitude (e.g. long-range) they
would append a convergence factor of the form $e^{-\a N^2}$ to the series
(\ref{reference_parition}). This strategy is being investigated.
However the non-next neighbor forces of physical interest are not necessarily
long-range, and have a substantial attractive component.
If the monomers are prevented from clustering, their forces would again
contribute to the order of $N$, and would simply modify the reference system
pathology.
Short range forces would reinforce this prohibition, but in the model we will
attend to, it is mainly the profile shaping due to the confined external
potential that is relied on.

\section{Diagrammatic Analysis}
We proceed to the fully interacting homopolymer ensemble.
In terms of $e^{-\b v(r,r')} = 1 + f(r,r')$, the $N^{th}$ term in
(\ref{reference_parition}) becomes
\beq \label{expansion}
\int\ldots\int\Pi'_{i,j}\left(1+f(r_i,r_j)\right)\z(r_1)w(r_1,r_2)\z(r_2)w(r_2,r_3)\ldots
  w(r_{N-1},r_N) dr_1\ldots dr_N,
\eeq
where $\Pi'$ is the product over $\frac{1}{2}N(N-1)$ distinct unordered pairs;
thus a typical term of the full expansion, in terms of multinomials on f, would
be diagrammatically like the one shown in Fig.\ref{expansion},
\begin{figure}[t] 
\begin{center}
\includegraphics[width=8cm]{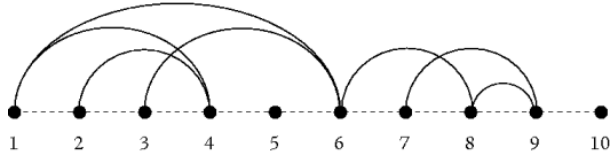} 
\caption{{\small Example of diagram in expansion (\ref{expansion}).}} \label{G-graphs}
\end{center}
\end{figure}
with multiplicative Boltzmann factor represented by:
\beqa
\CIRCLE &=& \z(r) \qquad{\mathrm{external\; field}} \nonumber \\
\begin{pspicture}(-.2,0)(3,.8)
\rput(.5,0.1){\rnode{A}{r\;}}
\rput(2.5,0.1){\rnode{B}{\;r'}}
\psset{linestyle=dashed}
\ncLine{A}{B}
\end{pspicture} &=& w(r,r') \qquad{\mathrm{next \; neighbor\; weight }}\nonumber \\
\begin{pspicture}(-.2,0)(3,.8)
\rput(.5,0.1){\rnode{A}{r\;}}
\rput(2.5,0.1){\rnode{B}{\;r'}}
\nccurve[ncurv=.8,linestyle=solid,angleA=65,angleB=120]{A}{B}
\end{pspicture} &=& f(r,r') \qquad{\mathrm{general\; pair \;Mayer \;factor}} \nonumber \\
\eeqa
All field points $r_i$ are integrated over.
It will be convenient to introduce the matrix:
\begin{pspicture}(-.2,0)(3,.2)
\rput(0.5,0.1){\rnode{A}{\Circle}}
\rput(2.5,0.1){\rnode{B}{\Circle}}
\psset{coilarm=.01,coilwidth=.1}
\nczigzag{A}{B}
\end{pspicture} 
denoted by $\Lambda$, for $N
\geq 2$, in which the endpoints have weight 1 and are not integrated over, but
the sum of all diagrams over all $N\geq 2$ is included.
Hence
\beq \label{partition_Lambda}
\Xi^{(p)}[\z] = \<1|\z\> + \<\z|\Lambda|\z\>
\eeq
where the vector $|\z\>=\z|1\>$.

Now various diagram reductions can be carried out.
The basic one stems from the observation that some of the vertices are
articulation points in the sense that their removal disconnects the diagram
into a left and right half, e.g. $r_6$ and $r_9$ in Fig.\ref{G-graphs}.

The end points are never given the status of articulation points.
If the matrix belonging to the sum of all \underline{connected} diagrams -
those with no articulation points, those with endpoints ($\Circle$) unintegrated, is denoted
by $C$, then clearly $\Lambda = C + C\,\z\,C+ C\,\z\,C\,\z\,C +\ldots$, or:
\beq \label{Lambda}
\Lambda = \left(I - C\,\z\right)^{-1} C,
\eeq
and (\ref{partition_Lambda}) can be written as
\beq \label{partition_C}
\Xi^{(p)}[\z]=\<1|\z\left((I-C\,\z\right)^{-1}|1\>.
\eeq
Of course, in the absence of the general pair interaction, $C=w$, and
(\ref{partition_C}) reduces to (\ref{reference_parition}).

\begin{figure}[t] 
\begin{center}
\includegraphics[width=8cm]{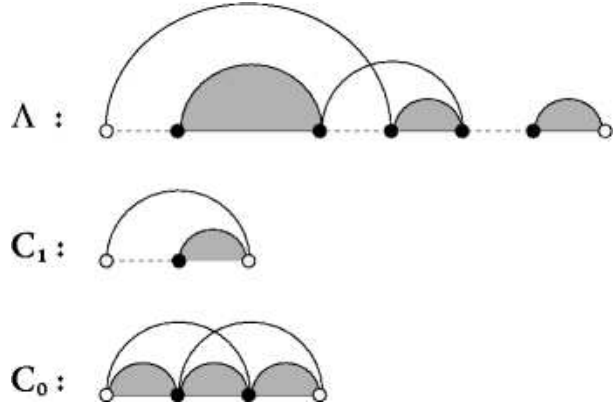} 
\caption{{\small Examples of graphs included in $\Lambda$,$C_1$, and $C_0$. The shaded gray areas
indicate the presence of any possible subgraph.}} \label{Lambda-fig}
\end{center}
\end{figure}

But C can also be built up from $\Lambda$.
To do this, observe that C consists of two types of diagrams, $C_1$, in which
the endpoints have a direct f-link to each other, and $C_0$ in which they
don't. See Fig.\ref{Lambda-fig} for examples.
Since the end-linked diagrams $\Lambda_1$ of $\Lambda$ are precisely the end-linked diagrams of
$C_1$ of $C$, and since $C_1=f*(\Lambda-\Lambda_1)$ we have
\beq \label{C1}
C_1 = f*(\Lambda - C_1),
\eeq
where $A*B$ denotes element by element multiplication of the matrices $A$ and
$B$.
On the other hand, $C_0$ is composed of $w$, together with all end-unlinked
diagrams that have no articulation points.
The latter are built by inserting $\Lambda's$ between the termini of any internal f
links, and so

\beq
C_0 = w\, +\,
\begin{pspicture}(-.2,0)(3.4,1)
\rput(0.2,0.1){\rnode{A}{\Circle}}
\rput(1.2,0.1){\rnode{B}{\CIRCLE}}
\rput(2.2,0.1){\rnode{C}{\CIRCLE}}
\rput(3.2,0.1){\rnode{D}{\Circle}}
\psset{coilarm=.01,coilwidth=.1}
\nczigzag{A}{D}
\nccurve[ncurv=0.8,linestyle=solid,angleA=60,angleB=120]{A}{C}
\nccurve[ncurv=0.8,linestyle=solid,angleA=60,angleB=120]{B}{D}
\end{pspicture} \,+\,
\begin{pspicture}(-.2,0)(4.4,1)
\rput(0.2,.1){\rnode{A}{\Circle}}
\rput(1.2,.1){\rnode{B}{\CIRCLE}}
\rput(2.2,.1){\rnode{C}{\CIRCLE}}
\rput(3.3,.1){\rnode{D}{\CIRCLE}}
\rput(4.2,.1){\rnode{E}{\Circle}}
\psset{coilarm=.01,coilwidth=.1}
\nczigzag{A}{E}
\nccurve[ncurv=0.8,linestyle=solid,angleA=60,angleB=120]{A}{C}
\nccurve[ncurv=0.8,linestyle=solid,angleA=60,angleB=120]{A}{D}
\nccurve[ncurv=0.8,linestyle=solid,angleA=60,angleB=120]{B}{E}
\end{pspicture}\,+\,\ldots
\eeq

But (\ref{C1}) can be written as
\beq
e*C_1 = f*\Lambda,
\eeq
where $e=1+f$, leading us to the second relation between $\Lambda$ and $C$:
\beq \label{e*C}
e*C = e*w + 
\begin{pspicture}(-.2,0)(1.4,1)
\rput(.2,.1){\rnode{A}{\Circle}}
\rput(1.2,.1){\rnode{B}{\Circle}}
\psset{coilarm=.01,coilwidth=.1}
\nczigzag{A}{B}
\nccurve[ncurv=1.2,linestyle=solid,angleA=60,angleB=120]{A}{B}
\end{pspicture}\,+\,
e *\begin{pspicture}(-.2,0)(3.4,1)
\rput(0.2,.1){\rnode{A}{\Circle}}
\rput(1.2,.1){\rnode{B}{\CIRCLE}}
\rput(2.2,.1){\rnode{C}{\CIRCLE}}
\rput(3.2,.1){\rnode{D}{\Circle}}
\psset{coilarm=.01,coilwidth=.1}
\nczigzag{A}{D}
\nccurve[ncurv=0.8,linestyle=solid,angleA=60,angleB=120]{A}{C}
\nccurve[ncurv=0.8,linestyle=solid,angleA=60,angleB=120]{B}{D}
\end{pspicture}\,+\,\ldots
\eeq
to accompany (\ref{Lambda}), now written as
\beq \label{closure}
\Lambda - C = C\,\z\,\Lambda
\eeq

\section{Mapping onto Fluids}
The practical theory of classical simple fluids in thermal equilibrium
has had a long period of development, entailing a sequence of more and more
sophisticated approaches \cite{Hansen}.
For uniform simple fluids not at thermodynamic singularities and not with long
range forces, little remains to be done. 
And the non-uniform situation is increasingly under control.
Let us indicate that we are speaking of a fluid - initially related to
the homopolymer under study only by having a common pair interaction,
$v_L(r,r') = v(r,r')$ - by appending a subscript $(L)$.
For present purposes, it's sufficient to observe that almost all of the older
approximations posit an algebraic relationship (or functional
relation in more recent versions) between the dimensionless pair correlation
function
\beq
h_L(r,r')=\frac{n_L(r,r')}{n_L(r)n_L(r')} -1
\eeq
and the direct correlation function $c_L(r,r')$, directly related to the
external potential - density linear response.

In classical thermal equilibrium, $h_l$ and $c_L$ satisfy identically the
Ornstein-Zernike equation \cite{OZ}, written in continuous matrix form as
\beq \label{OZ}
h_L - c_L = c_L n_L h_L,
\eeq
where $n_L$ is the particle density as diagonal operator
:$\<r|n_L|r'\>=n_L(r)\d(r-r')$.
We then need a second relation or closure to obtain both $h_L$ and $c_L$. 
One of the most effective reasonable approximations for short-range forces is
the P.Y. relation \cite{PY}, which in terms of Boltzmann factor $e$ and Mayer function $f=e-1$
of the pair interaction $v$ reads
\beq \label{PY}
e*c_L = f* h_L + f,
\eeq
where again the $(*)$ denotes element by element multiplication.

Comparison between the pair (\ref{OZ},\ref{PY}) and the pair
(\ref{closure},\ref{e*C}), is suggestive.
If we terminate (\ref{e*C}) at first order:
\beq
e*C = f*\Lambda + e*w,
\eeq
becomes identical with (\ref{OZ}) under the assumption
\beq \label{correspondance}
C = K c_L, \qquad \Lambda = K h_L, \qquad \z=n_L/K
\eeq
if the condition, for any fixed constant K,
\beq \label{K}
K f = e +w
\eeq
is satisfied - and (\ref{closure}) and (\ref{PY}) are identical as well.
Expressed in terms of the general pair interaction $v$ and the next neighbor
potential $\phi$, eq. (\ref{K}) says that
\beq \label{v_and_phi}
e^{\b v} + \frac{1}{K} e^{-\b\phi}=1
\eeq
is satisfied.
For example, it is sufficient to choose $v$ as, qualitatively, a reduced
version of the highly attractive localized $\phi$.
\begin{figure}[t] 
\begin{center}
\includegraphics[width=6cm]{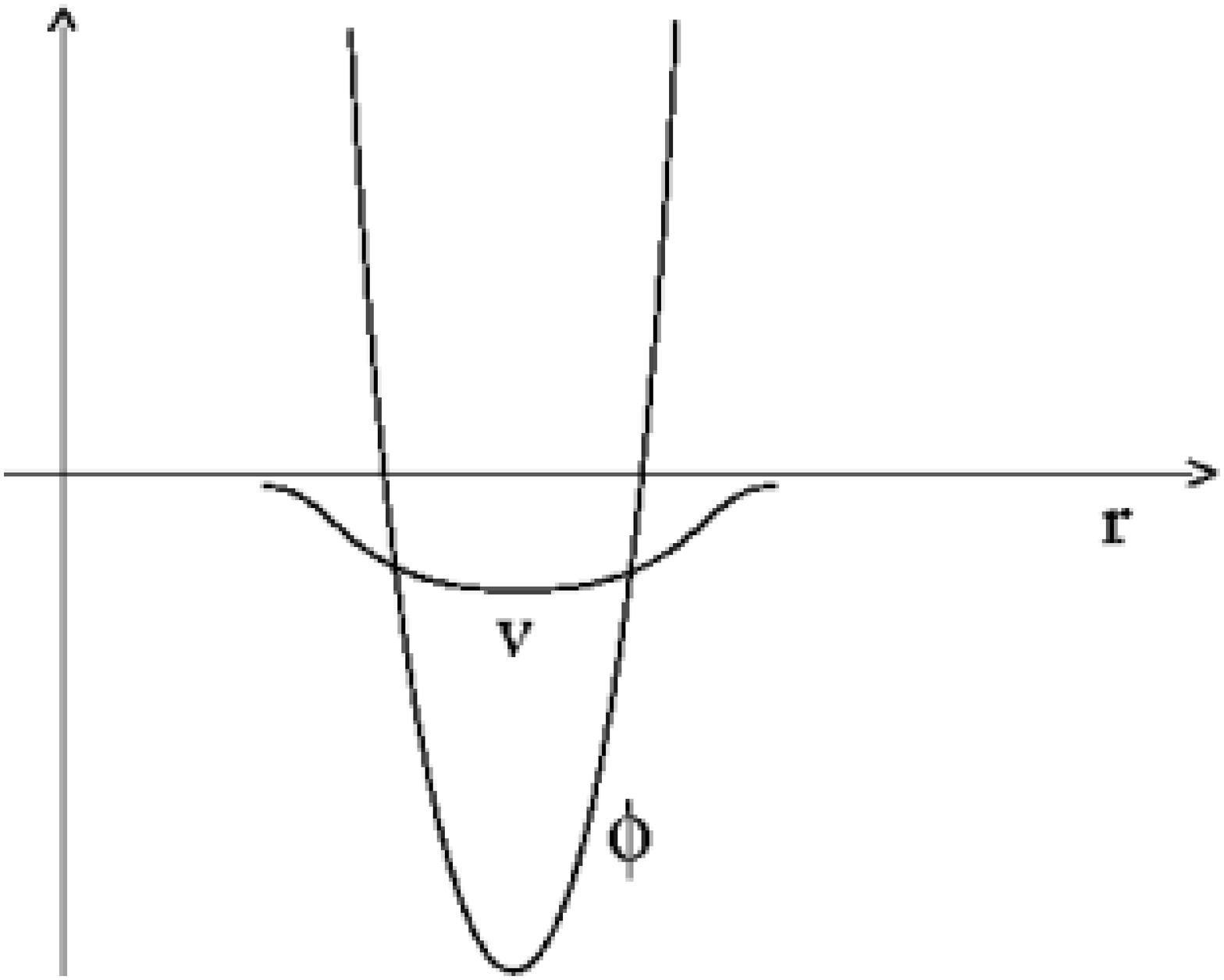} 
\caption{{\small Example of eq.(\ref{v_and_phi})}} \label{v_and_phi-fig}
\end{center}
\end{figure}
Note that according to our correspondence, $n_L= K \z$ may appear as a quite
high density in its confined domain.
Clearly, $(1/\b)\ln K + \phi >0$ is required by (\ref{v_and_phi}).

Once we have used the P.Y. approximation to produce a fluid corresponding to the
homopolymer under consideration, we can, in a logically tenuous fashion,
analyze the fluid by any other convenient approximation method.
Whatever approximation is used, the derived end result is the expression of
the polymeric monomer density resulting from the imposed containment potential represented by
$\z(r)$.
It's only necessary to translate this into equivalent fluid language.
From (\ref{density}), (\ref{partition_Lambda}) and  (\ref{correspondance}), we
have at once $\Xi^{(p)}=\<1|\z + \z\,\Lambda\,\z|1\> = \frac{1}{K}\<1|n_L+n_L h_L
n_L|1\>$, or
\beq
\Xi^{(p)}=\frac{\<1|S_L|1\>}{K},
\eeq
where $S_L(r,r')=n_L(r,r')-n_L(r) n_L(r')+n_L(r) \d(r-r')$ is the complete
Ursell function or generalized structure factor of the fluid, also
appearing as the density-density correlation function.

\section{Discussion}
The picture of an equivalent non-chain fluid is a bit deceptive.
At first glance, it resembles the familiar statistical model of biopolymers,
itself an offshot of the Lifschitz homopolymer condensation transition model \cite{Lifschitz}.
But the difference is seen most vividly in a situation in which there is an
unconstrained volume - constant external potential - bounded by a hard wall
container.
The ``dual'' fluid to the polymer would then have a constant density,
terminated by zero density at the boundary, and an internal potential structured to
produce the required uniform fluid.
Some meaningful conclusions can be arrived at in this situation without an
in-depth analysis.

Suppose we restrict attention to observations at the correlation length scale
of resolution.
On this scale, one knows that, clearly,
\beqa
S_L(r,r') &=& S_L(r-r';n_L(r)) \nonumber \\
&=& \d(r-r') \int S_L\left(R; n_L(r)\right) dR
\eeqa
where $S_L(R;n_L)$ is the density-density correlation for a uniform system of
density $n_L$.
Since
\beq
\int S_L(R;n_L) dR = \frac{\d n_L}{\d\b \mu_L},
\eeq
this implies that
\beq
\int\int S_L(r,r') dr dr' = \int \frac{\d n_L(r)}{\d\b\mu_L(r)} dr,
\eeq
and consequently that
\beqa \label{density_unif_fluid}
n(r) &=& \z(r) \frac{\d}{\d\z(r)}\ln\Xi^{(p)}[\z] \nonumber \\
&=& n_L(r) \frac{\d}{\d n_L(r)}\frac{\d n_L(r)}{\d\b\mu_L(r)}/\int
  \frac{\d n_L(r')}{\d\b\mu(r')} dr',
\eeqa
all of which is \underline{at} $n_L(r)=K \z_0$.
This would seem to indicate that the floppy chain, on averaging over all
configurations, would simply fill available space uniformly (since $n_L(r)$ has
no spatial dependence).
But this is only true if the correlation length is much smaller than the
diameter of the confining volume.
For example, in the primitive Van der Waals mean field model \cite{VdW}, in which
\beq
\ln n_L(r) = \b\mu_L(r) + \frac{1}{2}\int n_L(r) n_L(r')\left(e^{-\b
    v(r,r')-1}\right) dr',
\eeq
it is easy to see that (\ref{density_unif_fluid}) for a uniform fluid would
read
\beq
n = \frac{1}{V}\left( \frac{1 - 2 n_L^2 B}{1+2 n_L^2 B}\right),
\eeq
where $B$ is the second virial coefficient, negative for attractive $v$, and
$V$ the confinement volume.
Then, the needed divergence of $n\,V$ for large $V$ at "resonant" $n$, would
imply a divergent correlation length, and so in an exact solution, the density
would rise only gradually from its surface to bulk value, compatible with a
confined polymer transported throughout the volume.

One can of course also go beyond the equivalent fluid picture, but certainly
the domain untouched in our treatment is that of heteropolymers.
Statistical models are readily constructed by including species dependences in
the one and two-body potentials, but the statistical distribution is broad
unless the units are at least taken as short subsequences of monomers, thereby
introducing multiple internal degrees of freedom.
This strategy is being investigated, and in fact was implemented at a very
crude level some time ago in the context of including bond angle and dihedral
angle restrictions to accord with reality even in heteropolymers \cite{Muller}.

\section{Acknowlegdments}
S.P. would like to thank PCT-ESPCI and Volkswagenstiftung for current support,
and the New York University Department of Physics 
for the support during her Ph.D. when this work was mainly conducted.
The work of J.P. is partially sponsored by DOE, grant DE-FG02-02ER1592.

\end{document}